\documentclass[aps,prb,twocolumn,showpacs,floatfix]{revtex4}

\usepackage{dcolumn}
\usepackage[english]{babel}
\usepackage{amsfonts,amsmath,amssymb,bm}

\begin{document}

\title{Quantum LC - circuits with diffusive modification of the continuity equation}
\author{E.~Papp$^{1}$,~C.~Micu$^{2}$~and~O.~Borchin$^{3}$}
\email{erhardt_papp_2005@yahoo.com}
\thanks{$^{3}$E-mail:ovidiuborchin@yahoo.com}
\address{$^{1,3}$Department~of~Physics,~West~University~of~Timisoara, 300223, Romania}
\address{$^{2}$Department of Physics, North University of Baia Mare, 430122, Romania}%

\date{\today}

\begin{abstract}

Proofs are given that the quantum-mechanical description of the
LC-circuit with a time dependent external source can be readily
established by starting from a general discretization rule of the
electric charge. For this purpose one resorts to an arbitrary but
integer-dependent real function F(n) instead of n. This results in a
nontrivial generalization of the discrete time dependent
Schrödinger-equation established before via F(n)=n. Such
generalization leads to site-dependent hopping amplitudes as well as
to diffusive modification of the continuity equation. One shows, in
particular, that there are firm supports concerning rational
multiples of the elementary electric charge.


\end{abstract}

\maketitle

\section{Introduction}

\qquad Quantum transport of carriers in nanoscale systems has recived much
interest during the last two decades \cite{kram}-\cite{akker}. It has been
realized that fluctuations of the electric current are able to be implemented
by virtue of the discretness of electric charge Q \cite{lesov}-\cite{flor}.
Such Issues opened the way to the quantum-mechanical description of
$RLC$-circuits. Accordingly, current fluctuations have to be understood as
typical manifestations of appropiate quantum-mechanical Hamiltonians
incorporationg complementary charge and magnetic flux observables. Studies in
this field look promising, as they provide ideas for further technological
developments. The discretized charge reffered to above means that the
application of the discrete calculus looks rather suitable. The understanding
is that discrete tight binding models rely naturally on semicnductor quantum
wells and nano-electronic devices \cite{boyki}. The aim of this short paper is
to discuss in some more detail the quantum-mechanical description of the
mesoscopic $LC$-circuit with a time dependent voltage source $V_{s}(t)$. So
far the discrete Schrodinger equation characterizing the LC-circuit has been
established by starting from the charge eigenvalue equation \cite{lichen}%
,\cite{chendai}-\cite{lisun}:%
\begin{equation}
Q_{q}\left\vert n\right\rangle =nq_{e}\left\vert n\right\rangle \label{elect}%
\end{equation}
where $n$ is an integer playing the role of the discrete coordinate.
This shows that the electric charge gets quantized in units of the
elementary electric charge $q_{e}\equiv e$, as indicated by $(2.2)$
in \cite{lichen}, or by $(1)$ in \cite{chendai}-\cite{lisun}. One
could also say that $q_{e}=2e $ when dealing with Cooper-pairs
\cite{apenk}. However, more general charge quantization rules can
also be proposed. For this purpose we shall resort to an arbitrary
$n$-dependent function $F(n)$\ instead of $n$. This results in a
generalized counterpart of the discrete Schr\"{o}dinger-equation \
relying on (\ref{elect}) as well as in non-trivial modification of
the charge conservation law. Such modifications indicate that we
have to account for diffusion effects. One start from an appropriate
implementation of the canonically conjugated observable, i.e. of
suitable magnetic flux operators. For this purpose the charge
discreteness will be handled by\ applying $left$- and $right$-hand
discrete derivatives, i.e ${ \nabla}$ and ${ \Delta}$ \cite{nikif},
to charge eigenfunctions one deals with. This provides a pair of
non-Hermitian but conjugated magnetic flux operators. The product of
such operators is then responsible for the Hermitian operator of the
square magnetic flux. Of curse, the Hermitian magnetic flux
operator, which plays the role of the momentum, can also be readily
established in terms of a subsequent symmetrization.

\section{Preliminaries and notations}

\qquad We have to recall that the classical RLC-circuit is described by the
balance equation:%
\begin{equation}
L\frac{dI}{dt}+IR+\frac{Q}{C}=V_{s}\left(  t\right)  \ , \label{lrcform}%
\end{equation}
in accord with Kirchhoff's law, where the current is given by $I=dQ/dt$, as
usual. Inserting $R=0$, leads to the Hamiltonian%
\begin{equation}
\mathcal{H}_{c}\left(  Q,\frac{\Phi}{c}\right)  =\frac{\Phi^{2}}{2Lc^{2}%
}+\frac{Q^{2}}{2C}-QV_{s}\left(  t\right)  \ , \label{hamrz}%
\end{equation}
where $\Phi=ILc$ and $L$ stand for the magnetic flux and the inductance,
respectively. Indeed, (\ref{lrcform}) is produced by Hamiltonian equations of
motion characterizing (\ref{hamrz}):%
\begin{equation}
I=\frac{dQ}{dt}=\frac{\partial\mathcal{H}}{\partial\left(  \Phi/c\right)
}=\frac{\Phi}{Lc}\ , \label{ique}%
\end{equation}
and%
\begin{equation}
\frac{d}{dt}\left(  \frac{\Phi}{c}\right)  =-\frac{\partial\mathcal{H}%
}{\partial Q}=-\frac{Q}{C}+V_{s}\left(  t\right)  \ , \label{defhi}%
\end{equation}
as usual. This also means that the electric charge $Q$ and $\Phi/c$ are
canonically conjugated variables. This result suggest that the quantization of
the $LC$-circuit could be done in terms of the canonical commutation relation%
\begin{equation}
\left[  Q,\Phi\right]  =i\hbar c\ , \label{qiufi}%
\end{equation}
in which case one gets faced with the flux-operator \cite{flor}%
\begin{equation}
\Phi=-i\hbar c\frac{\partial}{\partial Q}\ . \label{fihace}%
\end{equation}

\qquad However, a such realization is questionable because the electric
charge, such as defined by (\ref{elect}) is not a continuous observable. This
means that the introduction of a discretized version of (\ref{fihace}) like%
\begin{equation}
\Phi_{q}=-i\frac{\hbar c}{q_{e}}{ \Delta} \label{phiqi}%
\end{equation}
for which $\Phi_{q}^{+}=-i\hbar c\nabla/q_{e}$ is in order. The Hermitian
time-dependent Hamiltonian of the quantum $LC$-circuit can then be established
as%
\begin{equation}
\mathcal{H}_{q}=\frac{\Phi_{q}^{+}\Phi_{q}}{2Lc^{2}}+\frac{Q_{q}^{2}}%
{2C}-Q_{q}V_{s}\left(  t\right)  \ . \label{hqfi}%
\end{equation}
in which $\mathcal{H}_{q}^{(0)}=\Phi_{q}^{+}\Phi_{q}/2Lc^{2}$ has the meaning
of the kinetic energy. The Hermitian momentum operator can also be readily
introduced as $P_{q}=\left(  \Phi_{q}^{+}+\Phi_{q}\right)  /2$. Note that
$right$- and $left$-hand discrete derivatives referred to above proceed as
\cite{nikif}%
\begin{equation}
{ \Delta}f\left(  n\right)  =f\left(  n+1\right)  -f\left( n\right)
\ , \label{discderleft}%
\end{equation}
and%
\begin{equation}
{ \nabla}f\left(  n\right)  =f\left(  n\right)  -f\left( n-1\right)
\ , \label{discderrih}%
\end{equation}
so that ${ \Delta}^{+}=-{ \nabla}$ and%
\begin{equation}
{ \nabla\Delta}={ \Delta}-{ \nabla}\ . \label{roule}%
\end{equation}

\qquad In addition, one has the product rule%
\begin{equation}
\nabla\left(  f\left(  n\right)  g\left(  n\right)  \right)  =g\left(
n\right)  \nabla f\left(  n\right)  +f\left(  n-1\right)  \nabla g\left(
n\right)  \ \label{prodroule}%
\end{equation}
and similarly for ${ \Delta}$.

\section{Generalized version of the electric charge quantization}

\qquad Looking for generalizations let us replace (\ref{elect}) by the charge
eigenvalue equation%
\begin{equation}
\widetilde{Q}_{q}\widetilde{\left\vert n\right\rangle }=q_{e}F\left(
n\right)  \widetilde{\left\vert n\right\rangle }\ , \label{elect1}%
\end{equation}
in which $F\left(  n\right)  $ is an arbitrary integer-dependent real
function. We have to assume that, in general, $\widetilde{\left\vert
n\right\rangle }$ is different from $\left\vert n\right\rangle $. Working
within the subspace spanned by $\widetilde{\left\vert n\right\rangle }$, one
finds%
\begin{equation}
\widetilde{Q}_{q}{ \Delta}=q_{e}F\left(  n+1\right)  { \Delta
}+q_{e}{ \Delta}F\left(  n\right)  \ , \label{eigenra}%
\end{equation}
and%
\begin{equation}
{ \nabla}\widetilde{Q}_{q}=q_{e}F\left(  n-1\right)  { \nabla
}+q_{e}{ \nabla}F\left(  n\right)  \ . \label{eigenle}%
\end{equation}

\qquad Performing the Hermitian conjugation gives ${ \nabla}%
\widetilde{Q}_{q}=q_{e}F\left(  n\right)  { \nabla}$ and ${ \Delta
}\widetilde{Q}_{q}=q_{e}F\left(  n\right)  { \Delta}$, where
$\widetilde{Q}_{q}^{+}=\widetilde{Q}_{q}$. Accordingly%
\begin{equation}
\left[  \widetilde{Q}_{q},{ \Delta}\right]  =q_{e}{ \Delta
}F\left(  n\right)  \left(  1+{ \Delta}\right)  \ , \label{qiuprod1}%
\end{equation}
and%
\begin{equation}
\left[  \widetilde{Q}_{q},{ \nabla}\right]  =q_{e}{ \nabla
}F\left(  n\right)  \left(  1-{ \nabla}\right)  \ . \label{qiuprod2}%
\end{equation}

\qquad Now we are ready to introduce rescaled magnetic flux operators like%
\begin{equation}
\widetilde{\Phi}_{q}=-\frac{i\hbar c}{q_{e}}\left(  \frac{1}{{
\Delta
}F\left(  n\right)  }{ \Delta}\right)  \ , \label{resmagflux}%
\end{equation}
which can be viewed as the generalized counterparts of (\ref{phiqi}) and%
\begin{equation}
\widetilde{\Phi}_{q}^{+}=-\frac{i\hbar c}{q_{e}}\left(  \frac{1}%
{{ \nabla}F\left(  n\right)  }{ \nabla}+\frac{1}{\Delta F\left(
n\right)  }-\frac{1}{\nabla F\left(  n\right)  }\right)  \ . \label{phiqi1}%
\end{equation}

\qquad Accordingly, the interaction-free Hamiltonian is given by%
\begin{equation}
\mathcal{H}_{q}^{\left(  0\right)  }\rightarrow\widetilde{\mathcal{H}}%
_{q}^{(0)}=\frac{\widetilde{\Phi}_{q}^{+}\widetilde{\Phi}_{q}}{2Lc^{2}%
}\ \label{interfreeham}%
\end{equation}
which can be rewritten equivalently as%
\begin{equation}
\widetilde{\mathcal{H}}_{q}^{\left(  0\right)  }=-\frac{\hbar^{2}}%
{2\widetilde{L}\left(  n\right)  q_{e}^{2}}(\widetilde{\Delta}-\nabla).
\label{interfreeham1}%
\end{equation}

\qquad This time the inductance gets rescaled as%
\begin{equation}
L\rightarrow\widetilde{L}\left(  n\right)  =L\left(  \nabla F\left(  n\right)
\right)  ^{2} \label{inducta}%
\end{equation}
whereas the discrete right hand derivative ${ \Delta}$ is replaced by%
\begin{equation}
\widetilde{{ \Delta}}=(1-G(n)){ \Delta}\text{ }. \label{disriha}%
\end{equation}
One has%
\begin{equation}
G\left(  n\right)  =1-\left[  \frac{{ \nabla}F\left(  n\right)
}{{ \Delta}F\left(  n\right)  }\right]  ^{2}. \label{geenef}%
\end{equation}
which leads to sensible effects for non-linear realizations of $F(n)$.Under
such conditions the discrete Schr\"{o}dinger equation implemented by the
generalized charge quantization condition (\ref{elect1}) is given by%
\[
\frac{\hbar^{2}(1-G(n))}{2\widetilde{L}\left(  n\right)  q_{e}^{2}}%
C_{n+1}\left(  t\right)  +i\hbar\frac{\partial}{\partial t}C_{n}\left(
t\right)  \ +\frac{\hbar^{2}}{2\widetilde{L}\left(  n\right)  q_{e}^{2}%
}C_{n-1}\left(  t\right)  =
\]
\qquad%
\begin{equation}
=\left[  \frac{\hbar^{2}}{\widetilde{L}\left(  n\right)  q_{e}^{2}}\left(
1-\frac{G\left(  n\right)  }{2}\right)  +\frac{q_{e}^{2}}{2C}F^{2}\left(
n\right)  -q_{e}F\left(  n\right)  V_{s}\left(  t\right)  \right]
C_{n}\left(  t\right)  \label{princip}%
\end{equation}
which leads to the usual result \cite{lichen}%

\begin{equation}
-\frac{\hbar^{2}}{2Lq_{e}^{2}}\left(  C_{n+1}+C_{n-1}\right)  +A=i\hbar
\frac{\partial}{\partial t}C_{n}\left(  t\right)  \label{princip1}%
\end{equation}
with $A=\left[  \frac{q_{e}^{2}}{2C}n^{2}-q_{e}nV_{s}\left(  t\right)
+\frac{\hbar^{2}}{Lq_{e}^{2}}\right]  C_{n}\left(  t\right)  $ via
$F(n)\rightarrow n$

\section{Modified charge conservation laws}

\ \qquad One seens that (\ref{princip}), which differs in a sensible manner
from (\ref{princip1}), has a rather complex structure such as involved by the
$n$-dependence of coefficients and especially of hopping amplitudes. However,
(\ref{princip}) as it stands provides useful insights for a more general
description of quantum mechanical circuits. Indeed, (\ref{princip}) produces a
modified continuity equation like%
\begin{equation}
\frac{\partial}{\partial t}\rho_{n}(t)+{ \Delta J}_{n}(t)=g_{n}(t)
\label{contieq}%
\end{equation}
where%
\begin{equation}
\rho_{n}(t)=\left\vert C_{n}(t)\right\vert ^{2} \label{chargdens}%
\end{equation}
denotes the usual charge density, whereas%
\begin{equation}
{ \Delta J}_{n}(t)=\frac{\hbar}{\widetilde{L}\left(  n\right)  q_{e}%
^{2}}\operatorname{Im}\left[  C_{n}(t)C_{n-1}^{\ast}(t)\right]
\label{relcurden}%
\end{equation}
stands for the related current density. The additional term in the continuity
equation is%
\begin{equation}
g_{n}(t)=G(n)\frac{\widetilde{L}\left(  n+1\right)  }{\widetilde{L}\left(
n\right)  }{ J}_{n+1}(t) \label{adterm}%
\end{equation}
which shows that there are additional effects, say diffusion processes, which
are able to affect the time dependence of the charge density. This results in
the onset of an extra charge density like%
\begin{equation}
\rho_{n}^{diff}(t)=-G(n)\frac{\widetilde{L}\left(  n+1\right)  }{\widetilde
{L}\left(  n\right)  }\int_{-\infty}^{t}{ J}_{n+1}(t^{\prime}%
)dt^{\prime} \label{diffcharde}%
\end{equation}
relying typically on the nonlinear attributes of the generalized charge
discretization function. The total charge density is then given by%

\begin{equation}
\rho_{n}^{tot}(t)=\rho_{n}(t)+\rho_{n}^{diff}(t) \label{totaldiffch}%
\end{equation}
in which it has been assumed that $\rho_{n}^{diff}(t)\rightarrow0$ when
$t\rightarrow\infty$.

\section{Other details}

\qquad After having been arrived at this stage, a systematic study of charge
discretization function would be in order, but such tasks go beyound the
immediate scope of this short paper. Choosing, however, a rational
generalization of (\ref{elect}) like%
\begin{equation}
F(n)=\frac{P}{Q}n \label{ralgen}%
\end{equation}
where $P$ and $Q$ are mutually prime integers, one finds immediately that
(\ref{princip1}) reproduces (\ref{princip}) just in terms of substitution%
\begin{equation}
q_{e}\rightarrow\widetilde{q}_{e}=\frac{P}{Q}q_{e} \label{qpeq}%
\end{equation}

\qquad This shows that $q_{e}$ and $Pq_{e}/Q$ can be placed on the same
footing. Accordingly, we are in a firm position to replace the elementary
electric charge $q_{e}$ by a rational multiple like $Pq_{e}/Q$, which
represents, strictly speaking, a non-trivial result. In addition,
(\ref{ralgen}) leads this time to $G(n)=0$, so that the usual form of the
charge conservation law gets restored. We have to recognize that nonlinear
realizations of the charge discretization function $F(n)$, although
interesting from the mathematical point of view, are not easily tractable.
Indeed, they lead, in general, to position dependent hopping amplitudes, to
anharmonic effects as well as to complex valued energy dispersion laws.
Moreover, in such cases the equivalence between the L-ring circuit and the
electron on the $1D$ lattice under the influence of the induced time dependent
electric field is lost and the same concerns inter-related dynamic
localization conditions \cite{chendai},\cite{dunken}. In other words, quantum
$LC$-circuits may be rather complex, but this is the point where
(\ref{princip}) looks promising for applications concerning diffusive motion
of electrons or of other carriers. Unusual commutation relations like%
\begin{equation}
\left[  \widetilde{Q}_{q},\widetilde{\Phi}_{q}\right]  =-i\hbar
c\left( 1+i\frac{q_{e}}{\hbar c}{
\Delta}F(n)\widetilde{\Phi}_{q}\right)
\label{commic}%
\end{equation}
and%
\begin{equation}
\left[  \widetilde{Q}_{q},\widetilde{\Phi}_{q}^{+}\right]  =-i\hbar c\left(
1-i\frac{q_{e}}{\hbar c}{ \nabla}F(n)\widetilde{\Phi}_{q}^{+}%
+\frac{{ \nabla}F(n)}{{ \Delta}F(n)}-1\right)  \label{commar}%
\end{equation}
have also to be mentioned. Such relationships can be viewed as non-Hermitian
versions of generalized canonical commutation relations acting on
non-commutative spaces \cite{conne},\cite{fadde}, which looks rather
challenging. Going back to (\ref{ralgen}) yields, however, a closed algebra
encompassing the kinetic energy, the momentum and the charge, as indicated
before \cite{lichen}.

\section{Conclusions}

\qquad In this paper we succeed to establish the quantum-mechanical
description of $LC$-circuits by starting from a rather general discretization
rule for the electric charge. To this aim one resorts to a real, but integer
dependent function $F(n)$ instead of $n$. This leads to the generalized
discrete Schr\"{o}dinger-equation (\ref{princip}), which reproduces the usual
result as soo as $F(n)=n$. A such generalized equation is able to incorporate
additional diffusion effects, as indicated by (\ref{contieq}) and
(\ref{diffcharde}). It is understood that such effects go beyond the charge
conservation proceeding usually in terms of ingoing and outgoing electron
flows. Equation (\ref{qpeq}) shows that we are in a firm position to replace
the elementary electric charge $q_{e}$ by a rational multiple like $Pq_{e}/Q$.
It is clear that proceeding via $C\rightarrow\infty$ leads to more general
descriptions of $L$-ring circuits, too. Selected realization of such
generalized descriptions deserve further attention, which looks promising fur
further applications.

\section{Acknowledgements}

\qquad The authors are indebted to CNCSIS/Bucharest for financial support.

\end{document}